\documentclass{elsart}
\usepackage{amssymb,epsfig,graphicx}
\begin{document}
\begin{frontmatter}

\title{Structural and Magnetic Phase Transitions in MnTe-MnSe solid solutions}

\author[gu]{Kapil E. Ingle}, \author[gu]{J. B. C. Efrem D'Sa}, \author[barc]{A. Das},
\author[gu]{K. R. Priolkar\corauthref{krp}}\ead{krp@unigoa.ac.in},
\corauth[krp]{Corresponding author}
\address[gu]{Department of Physics, Goa University, Goa, 403 206 India.}
\address[barc]{Solid State Physics Division, Bhabha Atomic Research Centre, Trombay, Mumbai 400 085 India}

\begin{abstract}
Neutron diffraction studies as a function of temperature on solid solutions of MnSe and MnTe in the Se rich region are presented. Interestingly as Te is doped in MnSe, the structural transformation to NiAs phase diminishes, both  in terms of \% fraction of compound as well as in terms of transition temperature. In MnTe$_{0.3}$Se$_{0.7}$, the NaCl to NiAs phase transformation occurs at about 40K and although it is present at room temperature in MnTe$_{0.5}$Se$_{0.5}$, its volume fraction is only about 10\% of the total volume of sample. The magnetic ordering temperature of the cubic phase decreases with increasing Te content while the hexagonal phase orders at the same temperature as in MnSe. Anomalies in thermal evolution of lattice parameters at magnetic ordering as well as structural transition temperatures indicate presence of magnetostructural coupling in these compounds.
\end{abstract}

\end{frontmatter}

\section{Introduction}
Studies of magneto-elastic coupling in intermetallic compounds have attracted considerable interest \cite{dung,wang,li}. Manganese telluride (MnTe) is a crossroad material between NiAs-type metallic transition-metal compounds and NaCl-type insulating manganese chalcogenides \cite{allen,bane,gos}. MnTe is a particularly interesting because it is a p-type semiconductor with a very high density of impurity charge carriers \cite{gos}. In spite of ordering antiferromagnetically with a Neel temperature, $T_N$ = 323K, it exhibits magnon drag effect \cite{was}. MnTe shows anomalies in transport and magnetic properties, like negative coefficient of resistance below 100 K and a sharp rise in susceptibility at around 83 K similar to a ferromagnetic transition which are closely related to its structural parameters \cite{efrem1}.

On the other hand, MnSe is an insulator with NaCl type structure at room temperature. Magnetic properties of MnSe show two antiferromagnetic transitions at about 265K and 130K respectively. Neutron diffraction studies have shown that MnSe undergoes a structural transition at about 270 K wherein a part of itself (~30\%) converts to NiAs phase which is antiferromagnetically ordered when it appears. While the high temperature cubic phase orders antiferromagnetically at 130K \cite{efrem2}. This partial transformation to NiAs phase can be attributed to presence of stacking faults in MnSe.

It is therefore of interest to explore the magnetic properties of solid solutions of MnTe and MnSe. Such attempts have been made in past which provide only a partial understanding of magnetic properties \cite{chehab,demi}. In light of new insights available on magnetic properties of MnTe and MnSe through neutron diffraction studies \cite{efrem1,efrem2} it would be interesting to revisit the solid solutions of the type MnTe$_x$Se$_{1-x}$, especially in the Se rich region and understand their magnetic properties.

In this paper we present detailed neutron diffraction studies on two compositions of MnTe - MnSe solid solutions. We have chosen MnTe$_{0.3}$Se$_{0.7}$ because 30\% of MnSe transforms to NiAs phase at 270K. The second composition chosen is MnTe$_{0.5}$Se$_{0.5}$ as it is on the NiAs to NaCl structural phase boundary in the phase diagram presented in Ref. \cite{chehab,demi}.

\section{Experimental}
The samples were prepared by mixing stoichiometric amounts of finely powdered Mn, Se, and Te and then pelletized and sealed in evacuated quartz ampoule below 10$^{-6}$ Torr. Subsequently these ampoules were slowly heated to 650$^\circ$C, annealed for 20 h and furnace-cooled. The samples were characterized by X-ray diffraction and were found to be single phase crystallizing in NaCl-type structure. Minor impurity peaks corresponding to NiAs type phase were noticed in MnTe$_{0.5}$Se$_{0.5}$. Magnetic susceptibility measurements were performed using an ac susceptometer in the temperature range 80 - 300 K in a field of 100 Oe. Neutron diffraction experiments were carried out in the temperature range 10 - 300 K and a wavelength of 1.24\AA~ using powder diffractometer at Dhruva, Trombay.

\section{Results}
Magnetic susceptibility of MnTe$_{0.3}$Se$_{0.7}$ and MnTe$_{0.5}$Se$_{0.5}$, measured as a function of temperature were very similar to those presented in \cite{demi} and showed broad transition with maximum at 130K and 120K respectively representing a transition from paramagnetic to antiferromagnetic state. Unlike in the case of MnSe \cite{efrem2}, no high temperature magnetic transition arising from the hexagonal NiAs phase was seen indicating a possibility of stable crystal structure in these two Te doped compounds. This is indeed interesting because the reported phase diagram shows a structural transformation from NaCl to NiAs type crystal structure up on Te doping. The transformation occurs close to 50\% doping \cite{chehab}. About 30\% of MnSe converts to antiferromagnetically ordered NiAs phase at around 270K \cite{efrem2} and therefore these doped systems should either have had NiAs type structure at room temperature or should have undergone a phase transformation from NaCl to NiAs phase at lower temperatures.

\begin{figure}[h]
\centering
\includegraphics[width=\columnwidth]{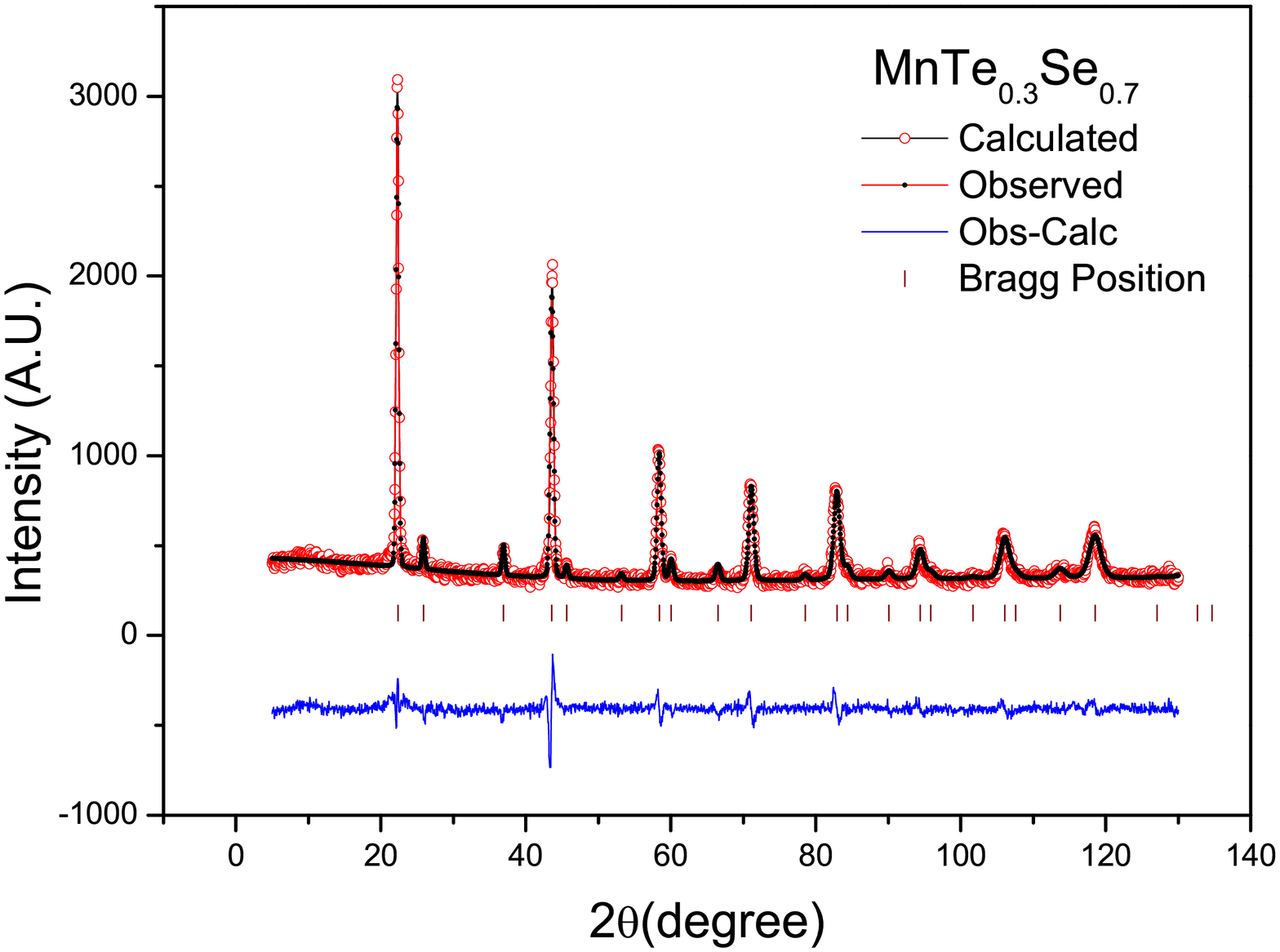}
\caption{\label{NDTe3Se7} Rietveld refined ND pattern of MnTe$_{0.3}$Se$_{0.7}$ at RT. The open circles show the observed counts and the continuous line passing through these counts is the calculated profile. The difference between the observed and calculated patterns is shown as a continuous line at the bottom of the two profiles. The calculated positions of the reflections are shown as vertical bars. The $R$-factors obatained were $R_P$ = 5.67, $R_{wp}$ = 8.36 and $R_{exp}$ = 5.13}
\end{figure}

\begin{figure}[h]
\centering
\includegraphics[width=\columnwidth]{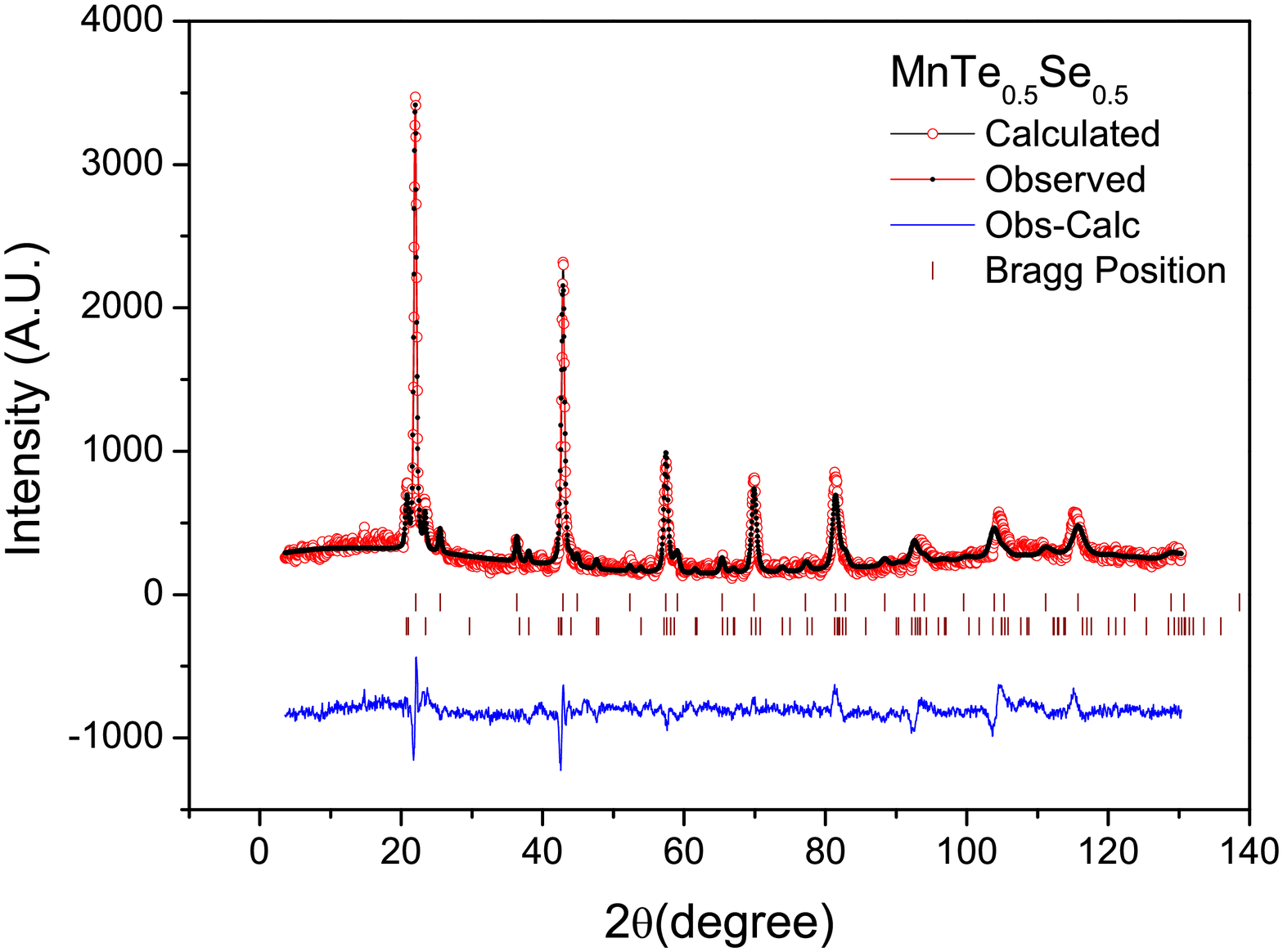}
\caption{\label{NDTe5Se5} Rietveld refined ND pattern of MnTe$_{0.5}$Se$_{0.5}$ at RT. The open circles show the observed counts and the continuous line passing through these counts is the calculated profile. The difference between the observed and calculated patterns is shown as a continuous line at the bottom of the two profiles. The calculated positions of the reflections are shown as vertical bars. The $R$-factors obtained were $R_P$ = 7.14, $R_{wp}$ = 10.31 and $R_{exp}$ = 5.79}
\end{figure}

In order to understand this rather surprising magnetic behavior, temperature dependent neutron diffraction measurements on the the two polycrystalline MnTe$_x$Se$_{1-x}$ (x = 0.3 and 0.5) have been carried out. Rietveld refinements of the diffraction data at various temperatures were carried out using the FULLPROF suite \cite{car}. The neutron diffraction patterns recorded in the angular range of $3^\circ \le 2\theta \le 130^\circ$ at room temperature (RT) for the two compounds are presented in Figs. \ref{NDTe3Se7} and \ref{NDTe5Se5} respectively.

MnTe$_{0.3}$Se$_{0.7}$ (Fig. \ref{NDTe3Se7}) presents a single phase NaCl type structure with lattice constant $a$ = 5.581 \AA. No additional peaks corresponding to hexagonal NiAs phase are visible. This is in agreement with the room temperature structure obtained from X-ray diffraction (not shown). In Fig. \ref{NDTe5Se5}, the diffraction pattern shows presence of a minor impurity phase in addition to major NaCl type cubic phase. The impurity phase can be refined with NiAs type hexagonal phase. The chemical composition of the hexagonal phase is similar to the nominal MnTe$_{0.5}$Se$_{0.5}$ and its relative composition is about 10\%. The presence of NiAs phase in the RT pattern is not surprising as this composition is close to the NaCl to NiAs transformation boundary its \% composition is much less. In MnSe, a partial conversion to NiAs phase occurs at 270K with about 30\% of the sample converting to hexagonal structure. Te doping should have facilitated formation of NiAs type phase. It may be mentioned here that amount of hexagonal phase does not change with lowering in temperature. Another point to be noted is that appearance of hexagonal phase in MnSe was also associated with its antiferromagnetic ordering. No peaks corresponding to antiferromagnetic ordering of the hexagonal phase are visible in the ND pattern recorded at RT.

\begin{figure}
\centering
\includegraphics[width=\columnwidth]{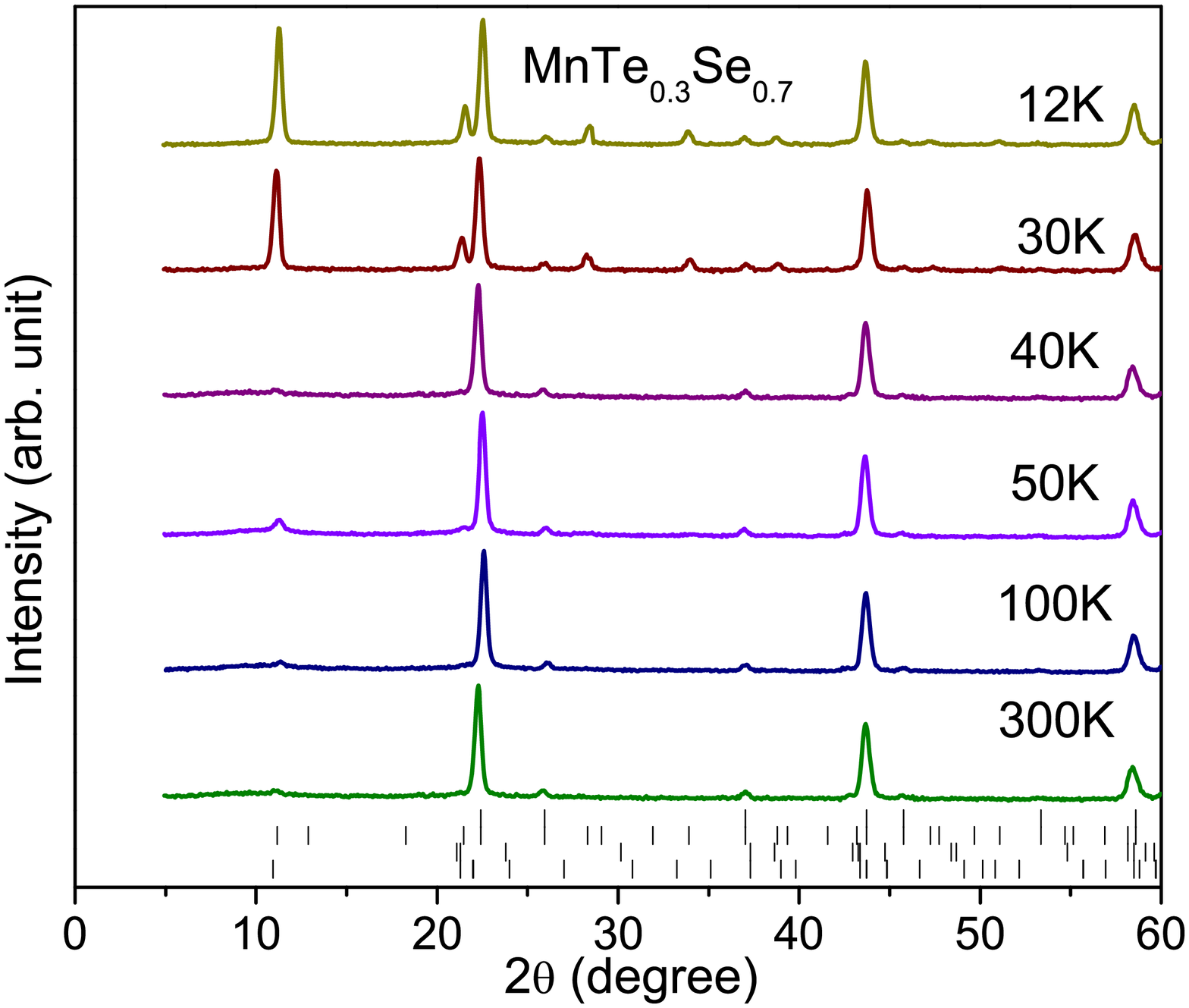}
\caption{\label{Te3Se7ndcompare} A comparison of neutron diffractions patterns of MnTe$_{0.3}$Se$_{0.7}$ recorded at various temperatures. The data is presented in limited 2$\theta$ range for clarity. Vertical bars shown at the bottom of the figure represent the positions of Bragg reflections calculated for cubic and hexagonal phases (nuclear and magnetic).}
\end{figure}

In MnTe$_{0.3}$Se$_{0.7}$, with lowering of temperature to about 120K, a weak peak at exactly  half the angle (2$\theta$ $\sim$ 11$^\circ$) of the most intense Bragg reflection emerges (Fig \ref{Te3Se7ndcompare}). This corresponds to antiferromagnetic ordering of the NaCl type parent phase with $T_N$ lying at about 130K. The appearance of magnetic (111) reflection suggests G-type antiferromagnetic order wherein Mn spins are aligned antiferromagnetically along the [111] direction. Similar spin alignment was also seen in MnSe. Between 100K and 50K the magnetic reflections grow in intensity indicating build up of Mn magnetic moment with decrease in temperature. However, with further lowering of temperature to 40K, the magnetic Bragg peak nearly disappears only to reemerge much more strongly at 30K. Additional reflections are also visible at this temperature and the diffraction pattern can only be refined by taking into account a small fraction of NiAs type hexagonal phase. This hexagonal phase is magnetically ordered at the temperature it appears with an antiparallel alignment of Mn spins along $c$ axis. This is again an interesting result and as will be discussed later, its origin lies in magnetostructural coupling.

\begin{figure}
\centering
\includegraphics[width=\columnwidth]{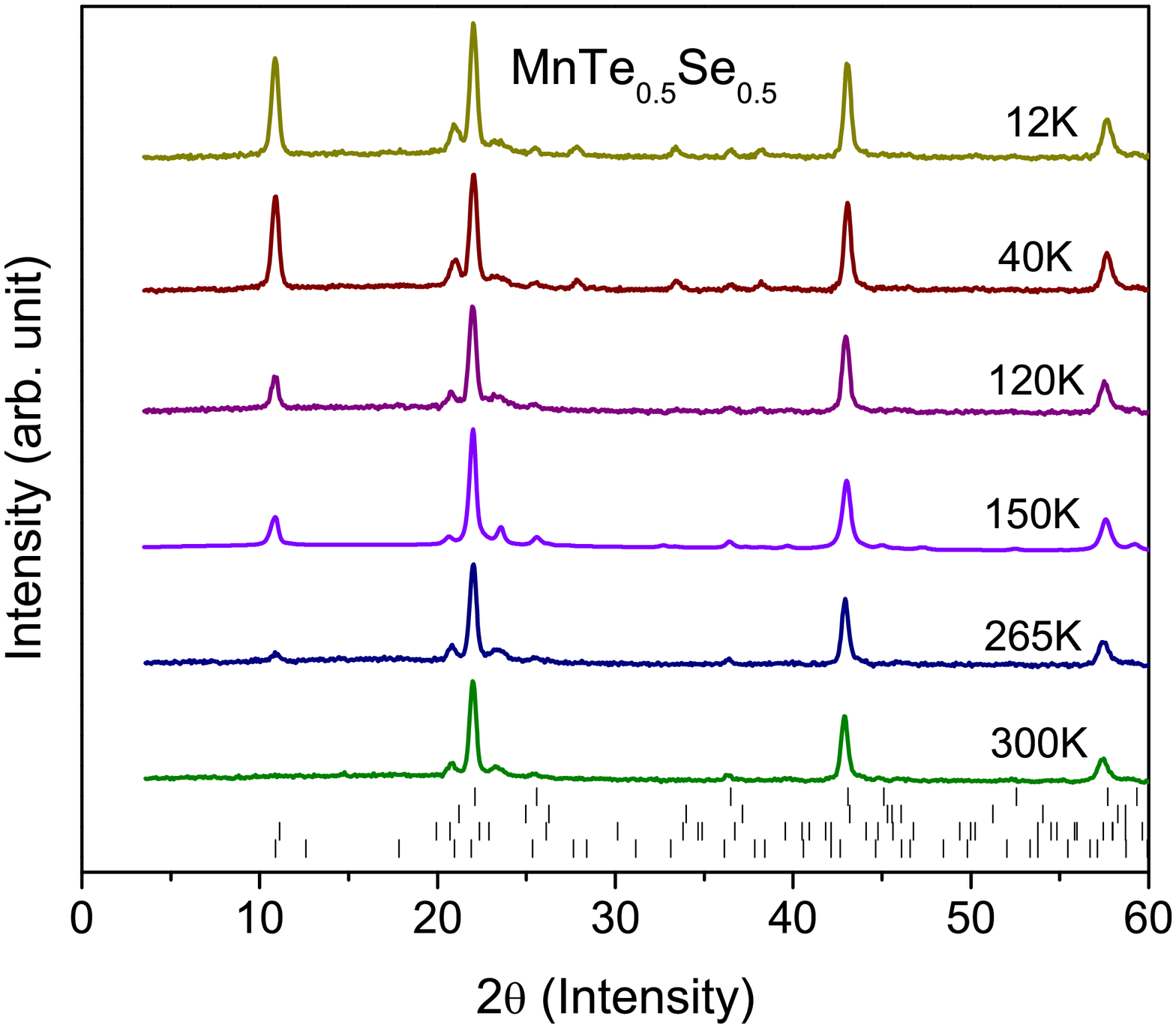}
\caption{\label{Te5Se5ndcompare} A comparison of neutron diffractions patterns of MnTe$_{0.5}$Se$_{0.5}$ recorded at various temperatures. The data is presented in limited 2$\theta$ range for clarity. Vertical bars presented at the bottom correspond to Bragg positions calculated for nuclear and magnetic contributions of cubic and magnetic phases.}
\end{figure}

The minor hexagonal phase present in MnTe$_{0.5}$Se$_{0.5}$ can be seen in Fig. \ref{Te5Se5ndcompare}, to order antiferromagnetically at about 270K. The hexagonal phase of MnSe also orders around the same temperature \cite{efrem2}. The major cubic phase orders antiferromagnetically at 100K. This is clearly evident from additional magnetic reflections that can be seen in diffraction patterns recorded below 265K and 100K respectively. There is no difference between the spin alignments of the two constituent phases in this compound and those found in parent MnSe suggesting thereby the mechanism of antiferromagnetic alignment to be same in these NaCl type Mn-chalcogenide semiconductors.

\begin{figure}
\centering
\includegraphics[width=\columnwidth]{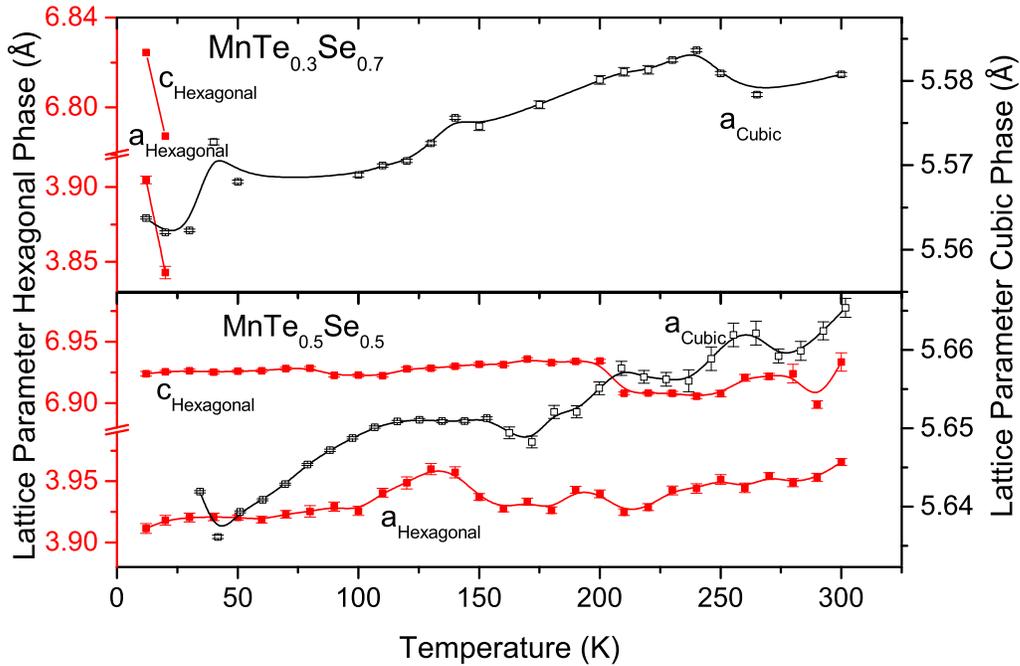}
\caption{\label{lattice} Evolution of lattice parameters of cubic and hexagonal phases as a function of temperature in MnTe$_{x}$Se$_{1-x}$, $x$ = 0.3 and 0.5.}
\end{figure}

The low temperature patterns were Rietveld refined using parameters obtained at RT as inputs. The parameters refined were cell parameters and magnetic moment parameters. The variation of cell parameters for each of the phases present in both the compounds is depicted in Fig. \ref{lattice}. In the case of MnTe$_{0.5}$Se$_{0.5}$ the lattice parameters of both cubic and hexagonal phases show nearly monotonic decrease with temperature. Anomalies seen around 270K and 110K can be identified with magnetic ordering of cubic and hexagonal phases. Like in case of MnSe, the hexagonal phase is identified with stacking faults along the $c$-axis \cite{and,jac}. These faults arise due to the incomplete structural transition of MnSe wherein only $\sim$ 30\% of the compound is converted to hexagonal phase below 270K. In the present case, about 10\% of NiAs type phase is present throughout the temperature range studied here. This hexagonal phase orders antiferromagnetically below 270K. The anomalies present in lattice parameters at the magnetic ordering temperatures hint towards a correlation between magnetic and structural degrees of freedom.

The neutron diffraction patterns of MnTe$_{0.3}$Se$_{0.7}$ provide an interesting trend. The compound exhibits purely cubic structure at RT and does not undergo any structural transition down to 40K. This is irrespective of the fact that 30\% of Se is replaced with Te and in MnSe almost about the same fraction of the sample converts to hexagonal phase below 270K. Furthermore, it must be mentioned that MnTe crystallizes in NiAs type hexagonal structure \cite{a}. However, the variation of lattice parameters exhibit very distinct anomalies. The cubic cell parameter shows a step like increase at about 270K which corresponds to the structural transition temperature in MnSe. A small kink like structure is visible at about 130K which corresponds to antiferromagnetic ordering temperature of the cubic phase. Further down, another strong anomaly is visible at about 40K which is just above the appearance of hexagonal phase. It is at this temperature, that magnetic ordering of the cubic phase was nearly destroyed (see Fig. \ref{Te3Se7ndcompare}). The variation of Mn magnetic moment presented below will make this point even more clear. Nevertheless, the intimate relationship between lattice parameters, and magnetic ordering indicates presence of magnetostructural coupling in both these compounds.

\begin{figure}
\centering
\includegraphics[width=\columnwidth]{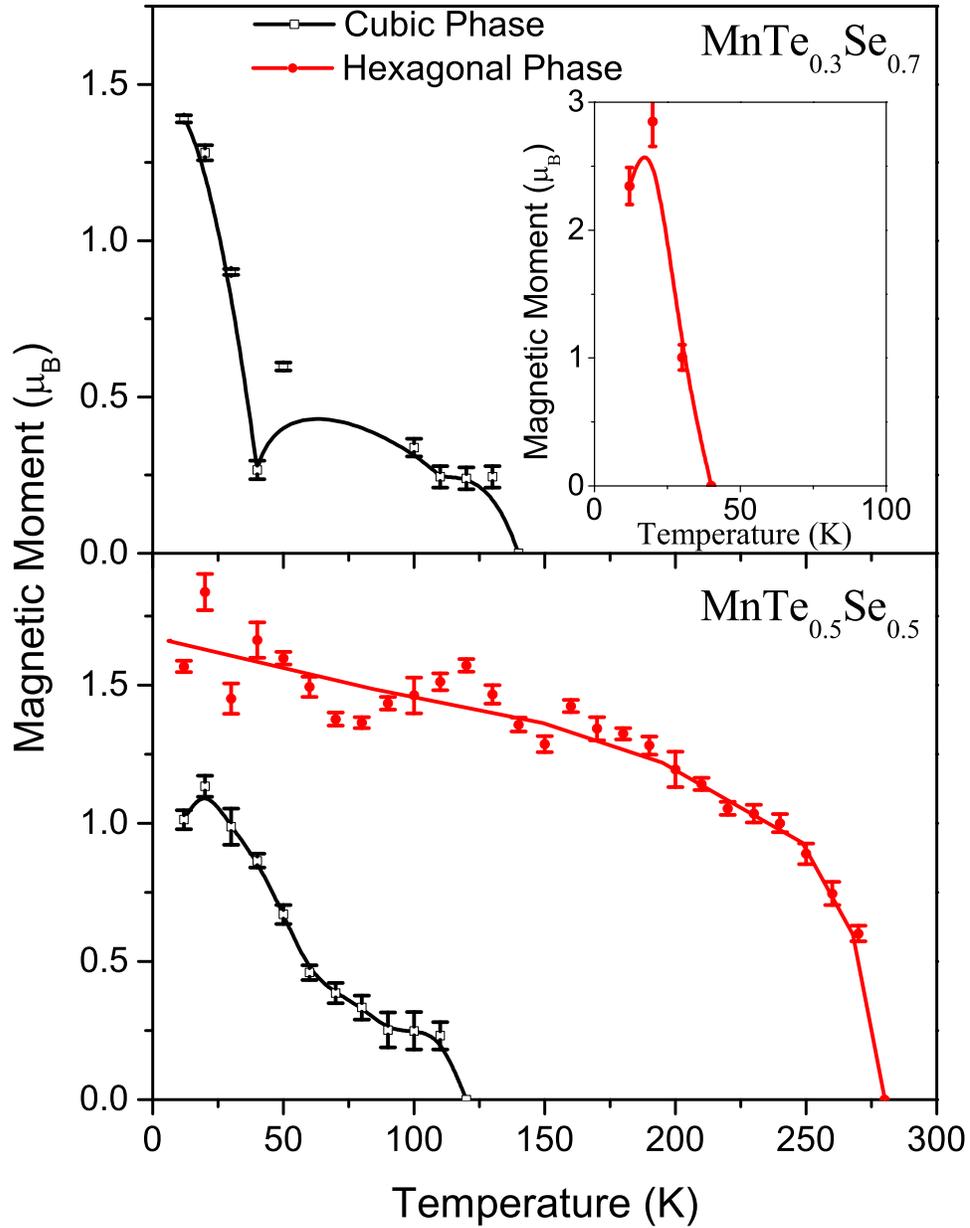}
\caption{\label{magmom} Variation of magnetic moment per Mn atom as a function of temperatures in MnTe$_{0.3}$Se$_{0.7}$ and MnTe$_{0.5}$Se$_{0.5}$.}
\end{figure}

Variation of magnetic moment per Mn ion in both, MnTe$_{0.3}$Se$_{0.7}$ and MnTe$_{0.5}$Se$_{0.5}$, obtained from Rietveld refinement of low temperature neutron diffraction data is presented in Fig. \ref{magmom}. In MnTe$_{0.5}$Se$_{0.5}$, the magnetic moment of both the phases increases with lowering of temperature. In case of hexagonal phase Mn ion exhibits near saturation below 100K and acquires a moment of about 1.5 $\mu_B$ at 12K. In the case of cubic phase however, the magnetic moment seems to continuously rise reaching about 1.0 $\mu_B$ at the lowest investigated temperature. In the other compound, the magnetic moment of the cubic phase remains quite low ($\sim$ 0.5 $\mu_B$) till about 50K below which it shows a sharp fall only to rise even more sharply at lower temperatures. Here the Mn ion acquires a maximum moment of 1.4 $\mu_B$ at the lowest temperature. Below 40K, there is also co-existing antiferromagnetically ordered hexagonal phase and the magnetic moment of Mn ions belonging to this phase show a sharp increase reaching a maximum value of about 2.5 $\mu_B$. A comparison of magnetic moment in MnSe and the present two compositions suggests an interesting correlation. In MnSe, the magnetic moment per Mn ion at 12K in both the phases was reported to be 3.3 $\mu_B$. With addition of Te in place of Se, the magnetic moment per Mn ion steadily decreases to 1.5 $\mu_B$ and 1.0 $\mu_B$ in hexagonal and cubic phases respectively in MnTe$_{0.5}$Se$_{0.5}$.

\section{Discussion}
MnTe$_{x}$Se$_{1-x}$ exhibits structural transition from NaCl cubic to NiAs hexagonal structure with increasing Te doping. Along with this structural transition the covalency between Mn $3d$ and chalcogen $p$ band is also affected giving rise unusual behavior of transport properties like magnon drag effect in MnTe \cite{a}. Although both MnSe and MnTe are antiferromagnets with high enough ordering temperatures, interplay between ferromagnetic and antiferromagnetic interactions has been reported in MnTe \cite{efrem1} and transition metal doped MnTe \cite{li}. The NiAs structure is a varient of hexagonal close packed structure and has a $c/a$ ratio very close to the ideal value of $\sqrt{8/3}$.

A hexagonal structure can be visualized in a NaCl unit cell with the lattice parameters satisfying a  relation, $a_{hexa} = a_{cubic}/\sqrt2$ and $c_{hexa} = \sqrt3 a_{cubic}/\sqrt2$. This gives a $c/a$ ratio of $\sqrt 3$ which is slightly higher than the ideal $c/a$ ratio normally satisfied by a NiAs structure. If we compare the lattice parameters of the two phases present in MnSe \cite{efrem2}, it can clearly be seen that $c$ parameter of the hexagonal phase is distinctly less than the one calculated from the cell parameter of NaCl cell. This mismatch in lattice parameters is perhaps the reason for the presence of stacking faults in NiAs type phase. In MnTe$_{0.5}$Se$_{0.5}$, hexagonal unit cell parameters are closer to the ones calculated from cubic cell constant over the entire temperature range and hence there is not much broadening of hexagonal peaks due to stacking faults. Interesting aspect however is that in spite of 50\% Te content, the amount of hexagonal phase is only about 10\%.

Surprising aspect however, is the near absence of hexagonal phase in MnTe$_{0.3}$Se$_{0.7}$. However, anomalies in lattice parameters are visible at 270K - magnetic ordering temperature of hexagonal phase in MnSe as well as MnTe$_{0.5}$Se$_{0.5}$. It is indeed surprising that hexagonal phase is not visible in this compound to temperatures down to 40K and could be related to the lattice parameter of the cubic phase. It is possible that c/a ratio of the hexagonal phase in this compound is greater than the ideal value of 1.63 and closer to $\sqrt 3$. In such a case the transformation of part of the compound to hexagonal phase will not be visible in presence of NaCl type cubic phase. With lowering of temperatures, the evolution of lattice parameters of the two phases changes and an antiferromagnetic NiAs type phase appears below 40K. The presence of competing cubic and hexagonal phases could be also the reason for a very small increase in magnetic moment of the cubic phase until the appearance of hexagonal phase. This also indicates the presence of magnetostructural coupling in this compound.

Antiferromagnetic ordering is prevalent in both the samples. The NiAs phase has type I ordering with Mn moments ferromagnetically aligned in the basal plane and these planes stacked antiferromagnetically along $c$ axis. The ordering temperature of the hexagonal phase in MnTe$_{0.5}$Se$_{0.5}$ is 270K which is nearly the same as in MnSe. In MnTe$_{0.3}$Se$_{0.7}$ the hexagonal phase is antiferromagnetic at the temperature it appears (T $\approx$ 30K). The parent cubic phase also orders antiferromagnetically  and the antiferromagnetic ordering temperature decreases with increasing Te content. This weakening of magnetic interactions in the cubic phase is linked to the structure. The expansion of cubic lattice parameter with addition of Te, facilitates the transformation from cubic to hexagonal structure at higher Te contents.

\section{Conclusions}

Crystal and magnetic structure of Se rich MnTe$_x$Se$_{1-x}$ have been studied using neutron diffraction. The sample with $x$ = 0.5 is close to R.T. structural phase transition boundary between NaCl type cubic to NiAs type hexagonal phase. NiAs phase is absent in MnTe$_{0.3}$Se$_{0.7}$ at higher temperatures and only appears at temperatures below 40K. This is because of the unique relationship between the lattice parameters of the cubic NaCl and hexagonal NiAs phases. Both the phases order antiferromagnetically. The magnetic ordering temperature of NiAs phase is 270K and is always magnetically ordered at T $<$ 270K it appears. The N$\acute{\rm e}$el temperature of the cubic phase slightly decreases with increasing Te content. The magnetic and structural transitions in these mixed chalcogenides are driven by a strong coupling between magnetic and structural degrees of freedom.

\section*{Acknowledgements}
Department of Science and Technology (DST), Government of India is acknowledged for financial support under the project No. SR/S2/CMP-57.


\end{document}